# Connection between dynamics and thermodynamics of liquids on the melting line

D. Fragiadakis and C.M. Roland

Naval Research Laboratory, Chemistry Division, Code 6126, Washington DC  20375-5342

ABSTRACT

The dynamics of a large number of liquids and polymers exhibit scaling properties characteristic of a simple repulsive inverse power law (IPL) potential, most notably the superpositioning of relaxation data as a function of the variable $TV^\gamma$, where $T$ is temperature, $V$ the specific volume, and $\gamma$ a material constant. A related scaling law, $T_m V_m^\Gamma$, with the same exponent $\Gamma=\gamma$, links the melting temperature $T_m$ and volume $V_m$ of the model IPL liquid; liquid dynamics is then invariant at the melting point. Motivated by a similar invariance of dynamics experimentally observed at transitions of liquid crystals, we determine dynamic and melting point scaling exponents $\gamma$ and $\Gamma$ for a large number of non-associating liquids. Rigid, spherical molecules containing no polar bonds have $\Gamma=\gamma$; consequently, the reduced relaxation time, viscosity and diffusion coefficient are each constant along the melting line. For other liquids $\gamma>\Gamma$ always; i.e., the dynamics is more sensitive to volume than is the melting point, and for these liquids the dynamics at the melting point slows down with increasing $T_m$ (that is, increasing pressure).

______________________________________________________________

## INTRODUCTION

When cooled or compressed, liquids undergo a series of transitions, associated with qualitative changes in dynamical properties [1]. At a relatively high temperature or low pressure, molecular cooperativity of the motions arises, evidenced by the onset of non-Arrhenius and non-Debye behavior. At lower temperature (or higher pressure) a dynamic crossover is observed, associated with a characteristic change in the temperature dependence of the relaxation time, along with deviations from the Stokes-Einstein and Einstein-Debye relations. Eventually, at a sufficiently low, pressure-dependent temperature, translational and rotational motions cease as the material becomes a glass. Studies of liquid dynamics as a function of temperature and pressure have shown that for a given liquid, the relaxation time is constant at each of these three dynamical transitions: the onset of cooperativity, the dynamic crossover, and the glass transition. That is, although the transition temperatures ($T_A$, $T_c$, and $T_g$) depend on pressure, the corresponding relaxation times at the transition are, to a very good approximation, independent of thermodynamic conditions [1].

For many liquids, these transitions cannot all be observed, usually due to crystallization – a genuine thermodynamic transition. For liquid crystals, which are rigid anisotropic molecules, the crystalline solid is reached via a series of transitions through mesomorphic (nematic, smectic, etc.) states, which exhibit orientational order but limited or non-existent long range translational order. Surprisingly, although the transition temperatures *per se* are pressure-dependent, the relaxation time for longitudinal reorientation (around the short axis of the molecule) of liquid crystals is constant at the transitions from mesomorphic states to the isotropic liquid and to the crystalline solid, as well as between different mesomorphic states [2,3]. This constancy of $\tau$ has thus far only been observed for mesogenic liquids. Whether the phenomenon extends to simple, isotropic liquids is unknown.

It is observed experimentally that for a wide variety of substances, including van der Waals liquids, polymers, and even ionic liquids and weakly hydrogen-bonded materials, relaxation times conform to the scaling law

$$\tau = f(TV^\gamma) \tag{1}$$

where $f$ is a function, $V$ the specific volume, and $\gamma$ a material constant [4,5]. If $\tau$ is constant at a transition, it follows that the temperature $T_c$ and volume $V_c$ at the transition are related by constancy of the product variable $T_c V_c^\Gamma$, where $\Gamma$ is a material constant. For a thermodynamic transition, such as for liquid crystals, it is not *a priori* obvious why the purely thermodynamic quantities $T_c$ and $V_c$, should be related by the exponent $\gamma$ which scales the *dynamics* of the system. However, the idea underlying density scaling is that the dynamic properties of viscous liquids are governed primarily by the repulsive component of the intermolecular potential, with changes in the long-range attractive part of the potential exerting a negligible effect. To the extent that this assumption is valid for the properties of interest, the potential can be approximated by an inverse power law (IPL) or "soft-sphere" intermolecular potential

$$U(r) \propto r^{-n} \qquad (2)$$

The IPL potential is scale-invariant: all dynamic properties and excess thermodynamic properties of the liquid, expressed in appropriately reduced units, are a function of the scaling variable $TV^\gamma$, with $\gamma=n/3$. In particular, the relaxation times, viscosities and diffusion coefficients (in reduced units; see below) conform to the scaling law (eq (1)) [6,7]. Thus, one can interpret the exponent $\gamma$, derived from scaling experimental relaxation times or viscosities, as providing a measure of the slope of the effective intermolecular potential.

For all state points for which $TV^\gamma$ is constant, not only are reduced dynamic quantities constant, but also the structure of the liquid. More generally, Dyre and coworkers proposed the term "isomorphs" for thermodynamic paths along which the Boltzmann probabilities of scaled molecular configurations are invariant [8,9], leading to the invariance of a multitude of dynamic properties. Liquids which have isomorphs (so-called strongly correlating liquids) also exhibit strong energy-pressure correlations [10,11], and, if the proportionality constant for energy and pressure fluctuations is state-point independent, density scaling. Liquids conforming to an IPL are strictly correlating, while those conforming to an IPL modified with an additional linear term are strongly correlating [10,12].

For an IPL liquid the temperature and volume at the melting point are related by the scaling law

$$T_m V_m^\Gamma = const. \qquad (3)$$

with $\Gamma=\gamma=n/3$; that is, the melting curve represents an isomorph [6]. From this follows the well-known Simon equation for the pressure-dependence of the melting temperature [13]

$$\frac{P-P_0}{a} = \left(\frac{T}{T_0}\right)^c - 1 \qquad (4)$$

where $T_0$ and $P_0$ are constants, and the parameter $c = (1+3/n)$ [14]. Similarly, reduced dynamical quantities for an IPL liquid are expected to be constant along the melting curve [15].

For a real liquid, the dynamics and the phase behavior are expected to be influenced by attractive interactions and other structure-sensitive terms (e.g., columbic forces). However, a large number of materials – in particular non-associating molecular liquids and polymers – exhibit characteristics associated with the IPL approximation, including density scaling of the dynamics.

Can eq. (3) be generalized to the melting point of such liquids? The fact that liquid crystal transitions, including those from a highly ordered, crystal-like mesophase such as the smectic E to isotropic liquid, conform to density scaling suggests that this might be the case. Dyre and coworkers [8] have predicted that along the melting curve eq (3) applies with $\gamma=\Gamma$, for liquids for which the IPL is a fair approximation. Such materials include non-associating liquids and liquid metals, but not, for example, ionic or hydrogen-bonded liquids.

We analyze literature data for a variety of non-associating liquids, in order to assess the extent to which properties relating to *thermodynamic* transitions, such as density scaling of the melting point (eq (3)), can be generalized from the idealized IPL to real liquids.

**RESULTS**

We analyze all liquids for which the following literature data were available: (a) the dynamics (viscosity, self-diffusion coefficient, or dielectric relaxation time) over a substantial range of temperatures and pressures, with sufficient accuracy to permit superpositioning in accord with eq.(1); (b) the equation of state (EoS) over the measurement range of the dynamic data and in the vicinity of the melting point; (c) the melting temperature as a function of pressure. For materials for which densities were measured together with viscosities or melting points, the former were used rather than an EoS from another source. For liquids having more than one crystalline form, we limited our analysis to the lowest-pressure crystal phase. Crystals with a degree of orientational disorder (for example, the low-pressure crystal phases of methane, neopentane, ethane, cyclohexane, carbon tetrachloride, and the odd-numbered *n*-alkanes with $n \geq 9$) show the same qualitative dependence of melting on $T$ and $V$ as orientationally ordered crystals; thus, they were included in this study.

**Density scaling of the dynamics.** In superpositioning data according to eq. (1), we use reduced dynamical quantities, defined using units of time $t_0 = v^{1/3}(kT/m)^{-1/2}$, length $l_0 = v^{1/3}$ and energy $E_0 = kT$, where *m* and *v* are molecular mass and volume. The reduced relaxation time, viscosity and diffusion coefficient are then

$$\tau^* = v^{-1/3}(kT/m)^{1/2}\tau$$
$$\eta^* = v^{2/3}(mkT)^{-1/2}\eta \qquad (5)$$
$$D^* = v^{-1/3}(kT/m)^{-1/2}D$$

The use of reduced units is motivated by the IPL potential, where density scaling applies to reduced quantities only [6,7,15]. The difference between the scaling of reduced and unreduced quantities is negligible in the supercooled regime; however, most of the liquids studied herein cannot be supercooled, so that the dynamics can only be analyzed over a relatively narrow range above the melting point. In such cases the difference between reduced and unreduced quantities can be substantial [7]; employing unreduced quantities either causes breakdown of the scaling or yields values of the exponent $\gamma$ which are unreasonably large or vary with the dynamic variable ($\tau$, $\eta$, or $D$).

To date, density scaling has been found to apply to a large number of non-associated liquids and polymers [4,5] and to molecular dynamics simulations based on inverse power law or Lennard-Jones type intermolecular potentials [9,10,11,12,16]. Here we show that it also holds for experimental viscosities and diffusion coefficients of simple monoatomic (argon, xenon, krypton) and diatomic (nitrogen, oxygen) liquids, as well as for a variety of other organic molecules for which scaling has not been previously reported. The obtained scaling exponents are listed in Table 1, with representative plots of viscosity versus $TV^\gamma$ displayed in Figures 1 and 2. For those liquids where more than one dynamic quantity was available, density scaling for each gave exponents equivalent within the experimental error.

**Density scaling of the melting point.** Figures 3 and 4 are double logarithmic plots of the melting temperature *vs.* melting volume for a representative subset of the materials studied. The data can be fit with straight lines. For certain materials, deviations from linearity were observed either at the highest or lowest pressures, presumably due to systematic errors in the EoS (which often must be extrapolated to the melting point using measurements at higher temperatures and/or lower pressures). From the linear fits we define a scaling exponent for the melting point

$$\Gamma = -\frac{d\ln T_m}{d\ln V_m} \qquad (6)$$

$\Gamma$ is analogous to the well-known thermodynamic potential parameter of liquid crystal transitions [2]. The range of the density and temperature data is quite limited, in most cases spanning less than one decade, as limited by the extent of the liquid phase in the $T$-$V$ domain. For this reason, we cannot infer unambiguously that $T_m V_m^\Gamma = const.$ from the data in figs. 3 and 4 alone. However, the parameter $\Gamma$ is a well-defined quantity for each liquid, describing the relative influence of temperature and volume on the melting point. In the case of the *n*-alkanes, the various pressure-dependent melting points were taken from different studies, performed using different techniques, with EoS state also obtained from different sources; nevertheless, the data in Fig. 5 show a systematic increase of the melting temperature and decrease of the slope (and thus $\Gamma$) with increasing chain length. An odd-even effect is also observed: odd-numbered *n*-alkanes tend to have higher scaling exponents than the values of even-numbered ones in a homologous series. A similar odd-even effect is well-known for melting temperatures, with even numbered *n*-alkanes tending to have the higher melting points [56].

The same procedure can also be applied to data from molecular dynamics simulations. Ahmed et al. [84] reported melting temperatures and volumes for a series of Lennard-Jones liquids with different repulsive exponents ranging from $m=7$ to 12. The data (Figure 6) are well described by the power law $T_m V_m^\Gamma = const.$ over the entire range studied, with the slope $\Gamma$ increasing with increasing steepness of the repulsive part of the intermolecular potential (i.e., larger $m$).

**Dynamic vs. thermodynamic scaling.** In Figure 7 we plot the scaling exponent for the dynamics, $\gamma$, versus the scaling exponent $\Gamma$ for the melting point. The diagonal, $\gamma = \Gamma$, corresponds to viscosity, relaxation time and diffusion coefficient, in reduced units, constant along the melting curve (i.e., this curve is an isomorph). For eight of the forty-three materials studied, $\gamma = \Gamma$ within the experimental uncertainty; these are argon, xenon, krypton, nitrogen, oxygen, methane and neopentane. For the other thirty-five materials $\gamma > \Gamma$, without exception.

The molecules of substances for which $\gamma = \Gamma$ are are roughly spherical, rigid, and lack polar bonds; this means their intermolecular potentials can be approximated as spherically symmetric. The presence of polar bonds (even if the dipoles cancel out so there is no net dipole moment) breaks the symmetry of the potential. Thus, although $\gamma = \Gamma$ for methane, $\gamma < \Gamma$ for carbon tetrachloride and tetrafluoride, which have the same molecular configuration as methane but polar C-F and C-Cl bonds. The same observation applies to isopentane ($\gamma=\Gamma$) compared tetramethylsilane ($\gamma<\Gamma$), where the C-C bonds of the former are replaced by (weakly) polar C-Si bonds.

Similarly, departure from a quasi-spherical shape results in unequal melting and dynamic scaling exponents. For example, in the case of the *n*-alkanes, the roughly spherical methane and ethane have $\gamma=\Gamma$, but for longer chains (having an elongated molecular shape and in intramolecular degrees of freedom) $\gamma<\Gamma$.

A parameter widely used to quantify the deviation of a molecule from sphericity is the acentric factor ω, introduced by Pitzer [85]. This factor, calculated from the critical properties of the liquid, is defined as $\omega = -\log(P_r) - 1$ at $T_r=0.7$, where $P_r$ and $T_r$ are the reduced vapor pressure and temperature relative to the critical point. A value of $\omega = 0$ corresponds to a perfectly spherical molecule. Acentric factors for the liquids studied are listed in Table 1; they vary from 0 to 0.8. In Figure 8, the difference $\gamma - \Gamma$ is plotted as a function of $\omega$. Only those liquids for which $\gamma=\Gamma$ have ω<0.1, with the exception of neopentane, for which ω=0.196. For the remaining liquids ω>0.1 and $\gamma>\Gamma$, although there is no correlation between the magnitude the acentric factor and the difference between the two scaling exponents.

We also note there is no correlation of the difference between $\gamma$ and $\Gamma$ with properties such as the molecular dipole moment, or the excess enthalpy or entropy of melting. The crystal structure also appears to be irrelevant: of the materials conforming to γ=Γ, argon krypton, xenon, methane and neopentane form face-centered cubic crystals, while the low pressure crystal phase of nitrogen is hexagonal, oxygen is monoclinic, and ethane is body-centered cubic. Moreover, the low pressure crystal phases of methane, neopentane, and ethane are plastic crystals, possessing some degree of orientational disorder. However, plastic crystallinity does not give rise to equivalent values for $\gamma$ and $\Gamma$: carbon tetrafluoride and carbon tetrachloride, tetramethylsilane, and certain odd-numbered alkanes have orientationally disordered crystals, but we find $\gamma>\Gamma$.

As stated, $\gamma$ is always greater than or equal to $\Gamma$; in no case do we observe $\Gamma>\gamma$. This implies that deviation from spherical symmetry of the intermolecular potential causes the dynamics to be more sensitive to volume than the melting point. This means that the effective intermolecular potential for dynamical properties is always steeper than that relevant to crystal melting. Whereas the dynamics reflect an orientational average of the intermolecular potential, the melting point is affected predominantly by the potential for molecules at specific angles, corresponding to their relative orientation in the crystal lattice. Moreover, as volume is reduced, molecular jamming increases the viscosity, with these strong steric interactions exacerbated for molecules having non-spherical shapes [86]. However, such effects do not obviously influence the relative stability of the crystal phase.

**DISCUSSION**

**Scaling exponents and molecular structure.** Collected in Table 1 are density scaling exponents for 43 liquids having relatively simple molecular structures. From these data we can draw inferences concerning the relationship between $\gamma$, which is governed by the steepness of the effective intermolecular potential, and structural properties. Polar liquids tend to have low values of $\gamma$, in agreement with both prior experimental results [87] and molecular dynamics simulations in which adding a permanent dipole to an asymmetric two-site Lennard-Jones molecule caused a similar decrease in $\gamma$ [9].

Increasing molecular flexibility, i.e., more internal degrees of freedom, corresponds to lower values of $\gamma$. For example, the *n*-akanes have smaller scaling exponents than the more rigid branched and cyclic alkanes. This reflects the softening of the interatomic potential due to the pressure-insensitive intramolecular bonds [87,88]. For the homologous series of *n*-alkanes having moderate (pentyl) or long chain lengths, both the dynamic and melting point scaling exponents continuously decrease with increasing chain length (Figure 5), with the difference between the two remaining approximately constant.

There are some cases for which a systematic change in chemical structure does not lead to corresponding changes in scaling exponents. For example, the presence of pendant methyl groups on benzene does not lead to systematic changes in $\gamma$, as seen by comparing data for toluene, xylenes, and mesitylene. In this respect the scaling behavior is similar to other properties (viscosity, glass transition temperature, nonexponentiality of relaxation, fragility, melting point and thermodynamic properties such as melting and other transitions), which likewise do not correlate in any obvious way with structure (that is, these fundamental properties cannot be determined *a priori* from knowledge of the chemical structure). Subtle effects related to packing and symmetries exert a nontrivial influence.

**Relationship to thermodynamic Grüneisen parameter.** According to the Lindemann criterion, melting transpires when the mean vibrational amplitude reaches a critical fraction of the interatomic distance [14]. Gilvarry [89] derived a form of the Lindemann equation (the so-called Lindemann-Gilvarry relation, widely used in geophysics [90]) that relates the change in volume along the melting curve to the thermodynamic Grüneisen parameter

$$\Gamma \equiv -\frac{d \ln T_m}{d \ln V_m} = 2\left(\gamma_G - \frac{1}{3}\right) \quad (7)$$

The dynamic scaling exponent can also be related to the Grüneisen parameter, based on the IPL approximation [91]

$$\gamma = 2\left(\gamma_G - \frac{1}{3}\right) \quad (8)$$

or from the assumption that the dynamics are governed by the excess entropy [92,93]

$$\gamma = \frac{C_V}{\Delta C_V}\gamma_G \quad (9)$$

A related formula is [12]

$$\gamma = \frac{C_V}{\Delta C_V}\left(\gamma_G - \frac{1}{3}\right) \simeq 2\left(\gamma_G - \frac{1}{3}\right) \quad (10)$$

Eqs. (7) - (10) indicate a link between the dynamics and the melting point that may extend more broadly to liquids for which an IPL potential is not applicable. However, since empirically $\gamma \sim \Gamma$ is observed only for a subset of the liquids in Table 1, modification of the Lindemann criterion or the arguments leading to eqs. (8) – (10) is necessary for consistency with experimental results for non-spherical liquids and those possessing internal degrees of freedom.

Molecular dynamics simulations have been used to study the effect of molecular shape on the scaling exponent. Using a repulsive exponent $m=12$, a density scaling exponent of $\gamma=5.0$ was obtained for a binary mixture of Lennard-Jones spheres [16]. For an asymmetric dumbbell molecule consisting of two different sized Lennard-Jones spheres (also with $m=12$), the scaling exponent increases to 6.1. And for the Lewis-Wahnström OTP model, consisting of three rigidly connected spheres, $\gamma=7.9$ [9]. Melting point data were not reported, but it would be of interest to determine $\Gamma$ for these asymmetric molecules. The expectation from the behavior of real materials is that $\Gamma<\gamma$.

**Viscosity at the melting point.** The dependence of liquid dynamics on pressure and temperature is important in the field of geophysics. Characterization of the relevant materials – iron and its alloys, silicates – is difficult because melting occurs at extreme temperatures and pressures of tens of GPa or more. Thus, extrapolations are required of properties measured for experimentally attainable conditions. One approach is to assume that the viscosity is constant at the melting point, independent of pressure. However, experimental data on geological materials indicate inferentially an increase of $\eta$ along the melting line with increasing $P$ [94,95]. This implies that at sufficiently high pressures, geophysical materials might vitrify at the melting point, suggesting that parts of the earth's core [94] or lower mantle [95] are in the glassy state. Of course, such conclusions are speculative, given the absence of direct measurements [96,97].

The extension of the results herein for molecular liquids to materials of geological interest could clarify this issue. Note that liquid metals are predicted to be strongly correlating [8], and accordingly density scaling should apply. Since metals are monoatomic, we anticipate that $\gamma=\Gamma$. If this is the case, dynamical quantities at the melting point are constant when expressed in reduced quantities, so that the viscosity of metals at the melting point can be expressed as

$$\eta(T_m) \propto V_m^{-2/3} T_m^{1/2} \tag{11}$$

This equation (with a universal prefactor) was proposed empirically by Andrade in 1931 [98] in the form [99]

$$\eta(T_m) = 5.1 \times 10^{-4} \frac{(AT_m)^{1/2}}{V_A^{2/3}} \tag{12}$$

where $A$ the atomic weight and $V_A$ the molar volume at the melting point. Eq. (12) has been used to extrapolate the viscosity at ambient pressure of various metals to the melting point. In liquids for which $\gamma>\Gamma$, the viscosity at the melting temperature would increase more strongly with pressure.

**Summary**

Two scaling relations, for the dynamics and for the melting behavior, yield respective exponents $\gamma$ and $\Gamma$. For spherical molecules lacking internal dipole moments, we find these exponents to be equivalent; however, more generally $\gamma > \Gamma$. The former condition prevails exactly for an IPL intermolecular potential, reflecting the appropriateness of the IPL approximation for many materials. From the scaling behavior two empirical relations, the Lindemann-Gilvarry and the Andrade equations, can be derived. In addition to providing insights regarding the forces between molecules, the scaling exponents can be used to estimate melting points at experimentally inaccessible temperatures and pressures.

**Acknowledgments**

This work was supported by the Office of Naval Research. We thank R. Casalini for useful discussions. D.F. acknowledges the National Research Council for a postdoctoral fellowship.

# Figure Captions

**Figure 1.** (Color online) Density scaling of the reduced viscosity for selected aliphatic and aromatic hydrocarbons.

**Figure 2.** (Color online) Density scaling of the reduced viscosity for liquid argon, krypton, nitrogen, carbon dioxide, trichloromethane and carbon tetrachloride.

**Figure 3.** (Color online) Melting temperature vs. melting volume for selected liquids from Table 1.

**Figure 4.** (Color online) Melting temperature vs. melting volume for the *n*-alkanes.

**Figure 5.** (Color online) Density scaling exponents for the melting point and for the dynamics, for the *n*-alkanes.

**Figure 6.** (Color online) Melting temperature vs. melting volume for the *m*-6 Lennard-Jones system (data from ref. [84]) The slope (scaling exponent for the melting point) increases with increasing steepness *m* of the repulsive part of the potential.

**Figure 7.** (Color online) Scaling exponent for the melting point versus for the dynamics, for the liquids in Table 1: nonpolar spherical molecules (circles), n-alkanes (squares), branched and cyclic alkanes (X), aromatic hydrocarbons (hexagons), halomethanes (diamonds), various other molecules: polar (triangles) and nonpolar (crosses).

**Figure 8.** (Color online) Difference between scaling exponents for the dynamics ($\gamma$) and the melting point ($\Gamma$) as a function of the acentric factor $\omega$. Symbols are as in Fig. 7.

**Figure 1.**

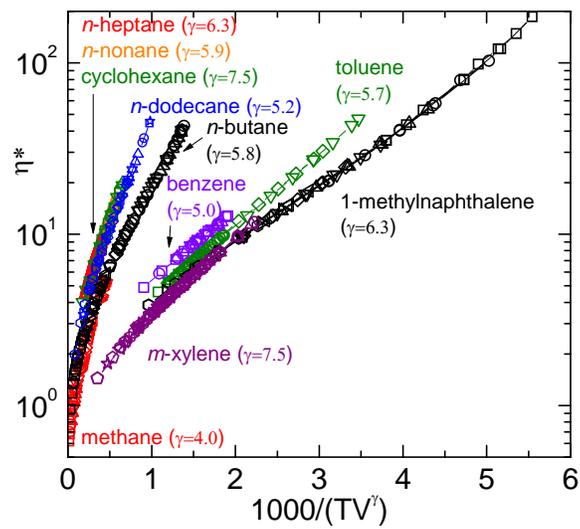

**Figure 2.**

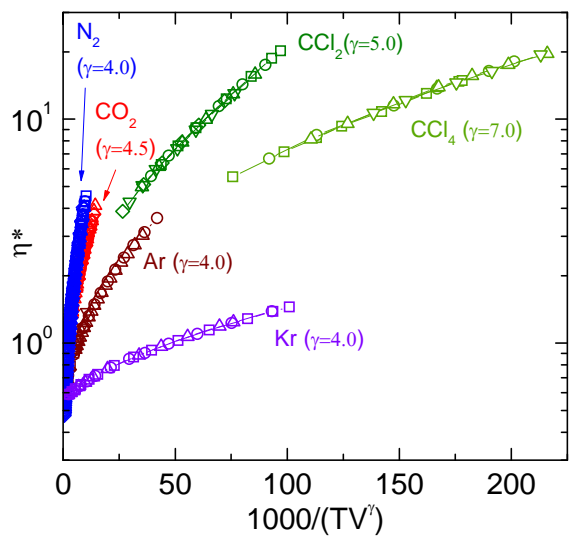

**Figure 3.**

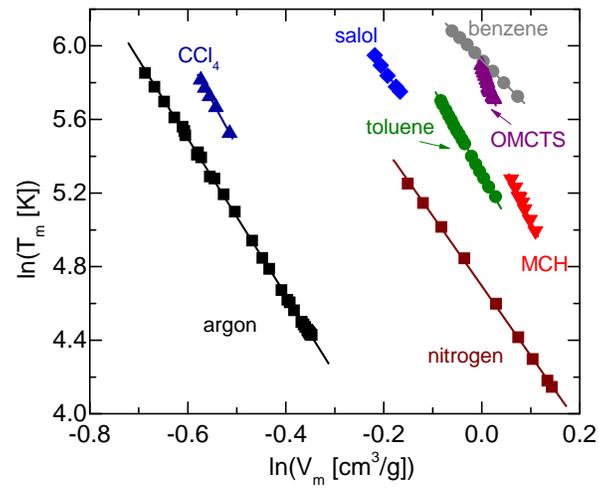

**Figure 4.**

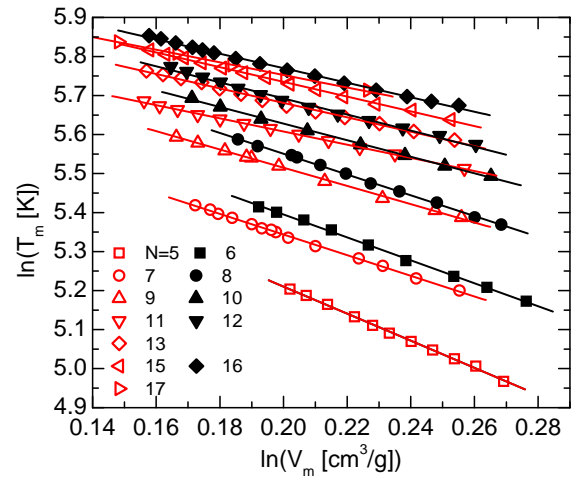

**Figure 5.**

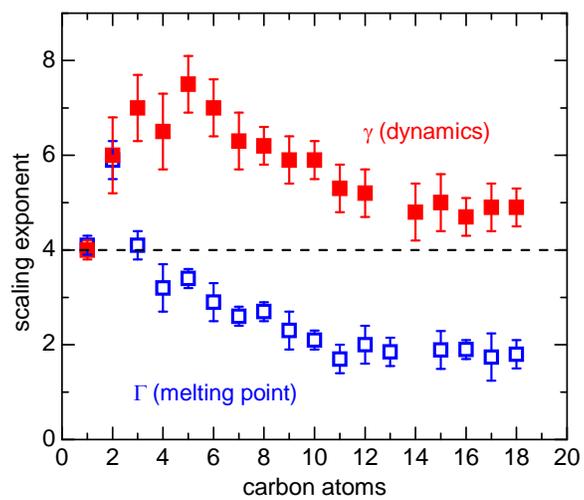

**Figure 6.**

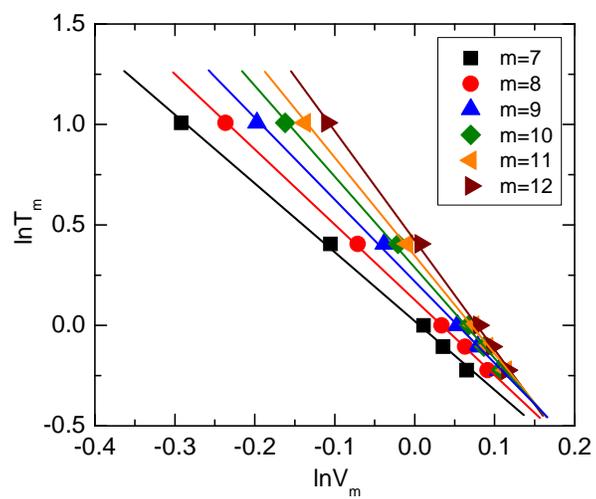

**Figure 7.**

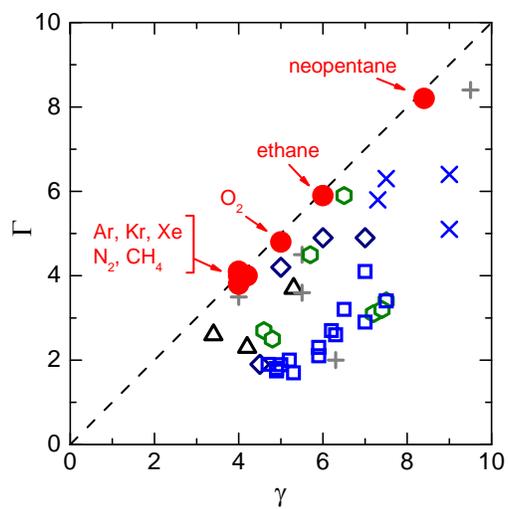

**Figure 8.**

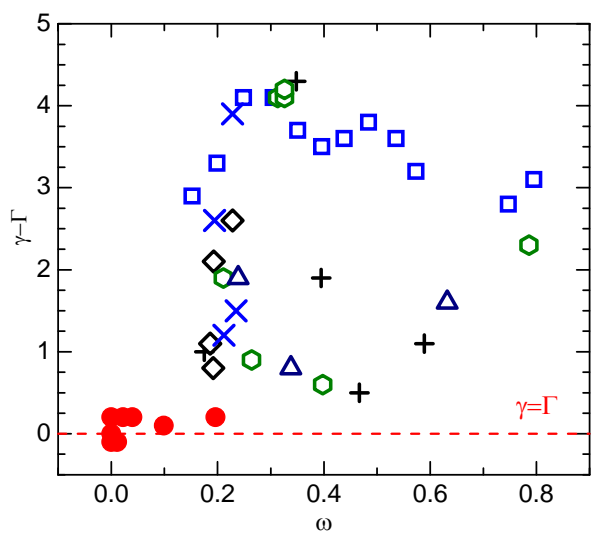

**Table 1.** Density scaling exponent for the melting point (Γ) and for dynamics (γ), acentric factor (from Ref. [17]), and literature sources for dynamics, melting points and equation of state for the liquids studied. Temperature ranges and maximum pressures for the dynamical data, as well as the range of melting points used in the scaling procedures are also given.

| No. | substance | Γ | γ | ω | T [K] | $P_{max}$ [MPa] | $T_m$ [K] | η or D or τ(T,P) | $T_m(P)$ | V(T,P) |
|---|---|---|---|---|---|---|---|---|---|---|
| 1 | argon | 4.0 | 4.0 | 0.000 | 90-500 | 400 | 84-348 | [18] (η) | [19, 20] | [21] |
| 2 | krypton | 4.0 | 4.0 | 0.000 | 298-348 | 200 | 116-157 | [21] (η) | [22] | [23] |
| 3 | xenon | 4.1 | 4.0 | 0.000 | 248-343 | 135 | 161-289 | [24] (D) | [25] | [18] |
| 4 | nitrogen | 3.8 | 4.0 | 0.040 | 285-473 | 2200 | 63-191 | [18] (η) | [26] | [26] |
| 5 | oxygen | 4.8 | 5.0 | 0.022 | 54-200 | 82 | 55-63 | [18] (η) | [27] | [27] |
| 6 | carbon dioxide | 1.9 | 4.5 | 0.228 | 220-300 | 453 | 216-265 | [28] (η) | [29] | [29] |
| 7 | dichloromethane | 4.2 | 5.0 | 0.192 | 186-306 | 200 | 177-200 | [30] (D) | [31] | [32] |
| 8 | carbon tetrachloride | 4.9 | 7.0 | 0.193 | 283-328 | 148 | 289-400 | [33] (D) | [34] | [32] |
| 9 | carbon tetrafluoride | 4.9 | 6.0 | 0.186 | 140-432 | 200 | 89-224 | [35] (D) | [36] | [18] |
| 10 | acetonitrile | 2.6 | 3.4 | 0.338 | 298-373 | 500 | 229-300 | [37] (η) | [38] | [39] |
| 11 | pyridine | 2.3 | 4.2 | 0.239 | 303-423 | 500 | 334-346 | [40] (D,η) | [41] | [39] |
| 12 | salol | 3.7 | 5.3 | 0.632 | 265-380 | 700 | 314-383 | [42] (τ) | [38] | [43] |
| 13 | octamethyltetrasiloxane | 8.4 | 9.5 | 0.589 | 293-493 | 200 | 289-363 | [44] (D) | [44] | [39] |
| 14 | hexafluorobenzene | 3.6 | 5.5 | 0.395 | 288-493 | 200 | 277-343 | [45] (D) | [46] | [32] |
| 15 | tetramethylsilane | 4.5 | 5.5 | 0.175 | 298-373 | 455 | 165-224 | [47] (η) | [48] | [32] |
| 16 | methane | 4.2 | 4.0 | 0.011 | 95-180 | 250 | 91-255 | [18] (η) | [49] | [49] |
| 17 | ethane | 5.9 | 5.0 | 0.099 | 136-454 | 200 | 90-185 | [50] (D) | [51] | [51] |
| 18 | propane | 4.1 | 7.0 | 0.152 | 112-453 | 200 | 86-125 | [50] (D) | [52] | [52] |
| 19 | n-butane | 3.2 | 5.8 | 0.199 | 135-400 | 69 | 113-125 | [18] (η) | [53] | [53] |
| 20 | n-pentane | 3.4 | 7.5 | 0.249 | 303-450 | 250 | 143-184 | [54] (η), [55] (η) | [56] | [57, 58] |
| 21 | n-hexane | 2.9 | 7.0 | 0.305 | 298-450 | 500 | 177-225 | [55] (η), [59] (η) | [56] | [57, 58] |
| 22 | n-heptane | 2.6 | 6.3 | 0.351 | 303-323 | 69 | 182-225 | [60] (η) | [56] | [57, 58] |
| 23 | n-octane | 2.7 | 6.2 | 0.396 | 283-473 | 375 | 216-269 | [61] (η), [62] (η) | [56] | [57, 58] |
| 24 | n-nonane | 2.3 | 5.9 | 0.438 | 303-323 | 69 | 220-268 | [60] (η) | [56] | [57, 58] |
| 25 | n-decane | 2.1 | 5.9 | 0.484 | 298-373 | 250 | 243-294 | [55] (η), [62] (η) | [56] | [57, 58] |
| 26 | n-undecane | 1.7 | 5.3 | 0.536 | 303-323 | 69 | 248-295 | [60] (η) | [63] | [64] |
| 27 | n-dodecane | 2.0 | 5.2 | 0.573 | 298-473 | 200 | 264-322 | [65] (η) | [63] | [64] |
| 28 | n-hexadecane | 1.9 | 4.7 | 0.747 | 298-373 | 425 | 291-348 | [59] (η) | [66] | [66] |
| 29 | n-octadecane | 1.8 | 4.9 | 0.795 | 323-473 | 90 | 301-353 | [65] (η) | [63] | [67] |
| 30 | neopentane | 8.2 | 8.4 | 0.196 | 267-450 | 120 | 257-390 | [68] (D) | [68] | [69] |
| 31 | isopentane | 5.1 | 9.0 | 0.228 | 298-328 | 200 | 113-151 | [70] (D) | [56] | [69] |
| 32 | cyclopentane | 6.4 | 9.0 | 0.194 | 298-328 | 200 | 178-251 | [70] (D) | [56] | [69] |
| 33 | cyclohexane | 6.3 | 7.5 | 0.212 | 313-383 | 214 | 279-417 | [71] (D,η) | [56] | [69] |
| 34 | methylcyclohexane | 5.8 | 7.3 | 0.235 | 298-363 | 500 | 147-196 | [72] (D), [73] (η) | [56] | [69] |
| 35 | benzene | 2.7 | 5.0 | 0.211 | 303-433 | 400 | 306-437 | [74] (η) | [75] | [76] |
| 36 | toluene | 4.5 | 5.7 | 0.264 | 298-363 | 149 | 177-900 | [61] (η), [73] (η) | [77] | [76] |
| 37 | o-xylene | 3.2 | 7.4 | 0.313 | 298-363 | 100 | 248-298 | [73] (η) | [38] | [76] |
| 38 | m-xylene | 3.4 | 7.5 | 0.326 | 298-473 | 200 | 228-283 | [62] (η), [73] (η) | [38] | [76] |
| 39 | p-xylene | 3.1 | 7.2 | 0.326 | 313-363 | 100 | 287-337 | [73] (η) | [38] | [76] |
| 40 | mesitylene | 5.9 | 6.5 | 0.398 | 298-313 | 280 | 221-306 | [78] (D) | [78] | [76] |
| 41 | 1-methylnaphthalene | 2.0 | 6.3 | 0.348 | 298-473 | 200 | 242-314 | [62] (η) | [75] | [76] |
| 42 | o-terphenyl | 3.5 | 4.0 | 0.467 | 256-308 | 125 | 329-477 | [79] (τ), [80] (τ) | [81] | [82] |
| 43 | n-dodecylbenzene | 2.5 | 4.8 | 0.786 | 298-363 | 100 | 269-292 | [73] (η) | [83] | [73] |